\documentclass{iopjournal}

\usepackage{amsmath,amssymb,graphicx,hyperref,multirow}

\usepackage[ignoreunlbld,norefs,nocites]{refcheck}

\usepackage[ascii]{inputenc}

\graphicspath{{./}}

\DeclareMathOperator\spn{span}

\begin{document}

\articletype{Paper}

\title{Baldereschi mean value points for three-dimensional Bravais
  lattices}

\author{N.\ D.\ Drummond$^1$\orcid{0000-0003-0128-9523}}

\affil{$^1$Department of Physics, Lancaster University, Lancaster LA1
  4YB, United Kingdom}

\email{n.drummond@lancaster.ac.uk}


\begin{abstract}
The Baldereschi point of a crystal is a wavevector in the Brillouin
zone at which every smooth periodic function of wavevector lies close
to its mean value. Although originally introduced in the context of
one-electron methods, mean-value points are ideal for explicitly
correlated many-electron methods such as quantum Monte Carlo
simulations, which can only use a single Bloch wavevector in the
Brillouin zone of a simulation supercell. We have therefore evaluated
and tabulated the Baldereschi mean-value points of all fourteen
three-dimensional Bravais lattices.
\end{abstract}

%\tableofcontents

\section{Introduction}

In most first-principles studies of condensed matter it is necessary
to integrate estimators of observable quantities over the first
Brillouin zone (BZ) of a crystal lattice in order to describe the
thermodynamic limit of infinite system size.  In single-electron
theories such as density functional theory (DFT), the BZ over which to
integrate is usually that of the primitive lattice. In many-electron
methods such as quantum Monte Carlo (QMC) algorithms, which use
explicitly correlated wave functions in supercells, the BZ over which
to integrate is that of the lattice of supercells.

Baldereschi \cite{Baldereschi_1973} introduced the idea of a
mean-value point, i.e., a single wavevector in the BZ such that the
mean value of any smoothly varying function of wavevector must lie
close to the value of the function at that point.  Mean value points
received some interest in the early days of electronic structure
calculations \cite{Evarestov_1983}, but were largely superseded in DFT
by the use of sets of special points \cite{Monkhorst_1976}, because in
DFT it is straightforward to use multiple Bloch wavevectors.  In QMC
calculations, on the other hand, each simulation can use only one
simulation-cell Bloch wavevector at a time \cite{Rajagopal_1994}.  We
can therefore either try to make that one wavevector as representative
of the average as possible \cite{Rajagopal_1994,Rajagopal_1995}, or we
can ``twist average'' by performing multiple QMC calculations with
different Bloch wavevectors to integrate over the supercell BZ
\cite{Lin_2001}.  In this paper we focus on the former approach.  If
one has to use a single wavevector in the BZ then that wavevector
should in general be the Baldereschi point.

Baldereschi points have previously been reported for the cubic family
of lattices \cite{Baldereschi_1973}, the two-dimensional (2D) lattices
\cite{Cunningham_1974}, and the hexagonal and tetragonal families of
lattices \cite{Bashenov_1977,Evarestov_1983}.  Recently a numerical
approach for evaluating Baldereschi points was made available
\cite{Stevanovic_2024}, although results were only presented for
some particular cases.  Here, we present a set of formulae for the
Baldereschi mean value points of all fourteen Bravais lattices in
three dimensions (3D)\@.

\section{Mean-value points in QMC calculations}

In QMC simulations of condensed matter we use supercells subject to
twisted periodic boundary conditions
\cite{Rajagopal_1994,Rajagopal_1995}.  Because QMC methods are
explicitly correlated in real space it is not possible to avoid the
use of a simulation supercell.  Translation of any electron through a
supercell lattice vector ${\bf R}_{\rm s}$ introduces a phase factor
$\exp(i{\bf k}_{\rm s} \cdot {\bf R}_{\rm s})$ into the many-body wave
function, where the ``twist'' ${\bf k}_{\rm s}$ is a simulation-cell
Bloch wavevector for the calculation.  If ${\bf k}_{\rm s}$ is the
$\Gamma$ point of the Brillouin zone (i.e., ${\bf k}_{\rm s}={\bf 0}$)
then we have pure periodic boundary conditions on the supercell.
Expectation values of energies and other quantities of interest are
symmetric functions of ${\bf k}_{\rm s}$ and, furthermore, are smooth
functions of ${\bf k}_{\rm s}$ in insulators, where the occupancy of
the single-particle orbitals in each Slater determinant does not
depend on the offset ${\bf k}_{\rm s}$ to the grid of the Bloch
wavevectors of the orbitals.  The only symmetry that can easily be
exploited within an individual QMC calculation is time-reversal
symmetry, which only results in a small efficiency saving.

Averaging QMC results in a finite supercell over Bloch wavevectors
${\bf k}_{\rm s}$ does not by itself yield the thermodynamic limit
because of long-range finite-size effects in both the periodic Coulomb
interaction \cite{Fraser_1996} and two-body correlations
\cite{Chiesa_2006}.  However, averaging over ${\bf k}_{\rm s}$ greatly
reduces quasirandom momentum quantisation effects, which are a very
important source of finite-size error in QMC calculations.  Since an
individual QMC calculation can only use a single ${\bf k}_{\rm s}$
value, the Baldereschi mean-value point is (by definition) the best
possible choice in the limit of slowly varying expectation values.

In metallic systems the ground-state occupancy of the single-particle
orbitals depends on the simulation-cell Bloch wavevector, leading to
discontinuities in properties such as energy per unit cell as a
function of ${\bf k}_{\rm s}$.  The most common solution to this
difficulty is twist averaging \cite{Lin_2001}, in which quantities of
interest are explicitly averaged over separate QMC calculations with
different ${\bf k}_{\rm s}$ values.  However, even in a metallic
system, the use of the Baldereschi point for the offset to a grid of
twists can reduce random fluctuations with system size
\cite{Drummond_2008,Azadi_2015}.

Where twist averaging is used, it is common practice to use just a
single twist ${\bf k}_{\rm s}$ to optimise the wave-function parameters.
The optimised parameters are then implicit functions of that twist, so
that the final twist-averaged results still depend on the twist ${\bf
  k}_{\rm s}$ that was used to optimise the wave function.  In the
limit of weak dependence of wave-function parameters on ${\bf k}_{\rm
  s}$, the Baldereschi mean-value point is by definition the point
that minimises the dependence of the final twist-averaged results on
the twist used for wave-function optimisation.

\section{Baldereschi mean value point}

\subsection{Symmetric functions of wavevector}

Consider a periodic crystal with Bravais lattice points $\{{\bf R}\}$.
Let $f({\bf k})$ be a smooth, periodic, integrable function of Bloch
wavevector ${\bf k}$.  Furthermore, suppose that $f({\bf k})$
transforms as the trivial one-dimensional irreducible representation
(irrep) $\Gamma_1$ of the point group of the Bravais lattice, i.e.,
$f({\bf k})$ is invariant under all symmetry operations of the
reciprocal lattice.  Then we may write the Fourier expansion of
$f({\bf k})$ as
\begin{equation} f({\bf k}) = \sum_{n=0}^\infty f_n
\sum_{{\bf R} \in \bigstar_n} \exp(i{\bf k}\cdot {\bf R}) = f_0 +
2\sum_{n=1}^\infty f_n \sum_{{\bf R} \in \bigstar^+_n} \cos({\bf
  k}\cdot {\bf R}) \equiv f_0 + \sum_{n=1}^\infty f_n A_n({\bf
  k}), \end{equation} where $\bigstar_n$ is the $n$th star of lattice
points that are mapped into each other by point symmetry operations,
arranged in ascending order of $|{\bf R}|$.  The first star simply
contains ${\bf R}={\bf 0}$ and the corresponding Fourier coefficient
$f_0$ is the average of $f({\bf k})$ over wavevectors ${\bf k}$ in the
first BZ\@.  By the inversion symmetry of Bravais lattices, all other
stars consist of $\pm {\bf R}$ pairs of lattice points, and hence we
can replace the sum of complex exponentials over $\bigstar_n$ with a
sum of cosines over $\bigstar^+_n$, where $\bigstar^+_n$ only contains
one out of each $\pm{\bf R}$ pair. We refer to $A_n({\bf k}) \equiv
\sum_{{\bf R} \in \bigstar_n} \exp(i{\bf k}\cdot {\bf R}) =
2\sum_{{\bf R} \in \bigstar^+_n} \cos({\bf k}\cdot {\bf R})$ as the
$n$th star function.  These star functions form an orthogonal basis
for symmetric functions $f({\bf k})$.

Observable quantities of physical interest, such as the total energy
or electronic charge density, are generally symmetric functions of
${\bf k}$ and, in any case, only the totally symmetric part of a
function of ${\bf k}$ contributes to its mean value
\cite{Baldereschi_1973}, so that there is no loss of generality
involved in considering only totally symmetric functions $f({\bf k})$.

To motivate the definition of the Baldereschi point, since we are
interested in situations in which $f({\bf k})$ is unknown ahead of
time (and indeed we may be simultaneously interested in many different
observable quantities), we consider an ensemble of random functions
$f({\bf k})$ with the symmetry of the reciprocal lattice and rapidly
convergent Fourier series.  In particular, suppose that for $n>0$,
each $f_n$ is drawn independently from a distribution whose mean is
zero and whose standard deviation $\sigma_n = s_0 \exp(-\lambda n)$
falls off exponentially with $n$, where $s_0$ and $\lambda$ are
positive constants.  Then the mean squared difference between $f({\bf
  k})$ and $f_0$ is
\begin{equation} \left< [f({\bf k})-f_0]^2 \right>_f 
= \sum_{n=1}^\infty s_0^2 e^{-2\lambda n} A_n^2({\bf
  k}), \label{eq:mse} \end{equation} where the angled brackets denote
the mean over the ensemble of functions $f$.

If $\lambda$ is sufficiently large then the mean squared difference is
exponentially dominated by the first term in equation (\ref{eq:mse}),
and so to minimise the mean squared difference at large $\lambda$ we
require ${\bf k}$ to satisfy $A_1({\bf k})=0$ [which may have the
  effect of determining some subsequent star functions $A_n({\bf k})$
  with $n>1$].  We must then find the first star function
$A_{t_2}({\bf k})$ that can be varied on the surface $A_1({\bf k})=0$
and either set $A_{t_2}({\bf k})$ to zero or minimise $A_{t_2}^2({\bf
  k})$ on the surface $A_1({\bf k})=0$ [which may have the effect of
  determining some subsequent star functions $A_n({\bf k})$ with
  $n>t_2$].  If a subsequent star function $A_{t_3}$ can be varied
subject to the conditions $A_1({\bf k})=A_{t_2}({\bf k})=0$ then we
must either set $A_{t_3}({\bf k})$ to zero or minimise $A_{t_3}^2({\bf
  k})$ subject to $A_1({\bf k})=A_{t_2}({\bf k})=0$.  The value of
${\bf k}$ thus determined, which minimises the mean squared difference
between $f({\bf k})$ and $f_0$ for sufficiently large $\lambda$, is
the Baldereschi point \cite{Baldereschi_1973}.

The Baldereschi point is purely determined by the Bravais lattice via
the star functions $A_n({\bf k})$; it does not depend on the
particular observable $f({\bf k})$ that is to be averaged over the
Brillouin zone.  For \textit{any} function $f({\bf k})$ that has the
symmetry of the reciprocal lattice and a sufficiently rapidly
convergent Fourier series, we expect its value at the Baldereschi
point to be close to its mean value over the Brillouin zone $f_0$;
furthermore the Baldereschi point is the point at which symmetric
functions with rapidly convergent Fourier series are on average
closest to their mean values.

For functions $f({\bf k})$ with slowly convergent Fourier series, the
Baldereschi point remains as good a choice as any if a single point in
the Brillouin zone is required to represent the average of $f({\bf
  k})$ over the Brillouin zone and the precise form of $f({\bf k})$ is
unknown ahead of time; in this case, however, the effectiveness of the
Baldereschi point is something to be explored empirically on a
case-by-case basis: see section \ref{sec:examples}.

\subsection{Stars of lattice points}

Each star of lattice points $\bigstar_n$ carries a
$\max\{\dim(\spn(\bigstar_n)),1\}$-dimensional representation of the
point group of the Bravais lattice, with matrices defined by the
symmetry transformations between the lattice points in the star.  Let
$r_n$ be the magnitude of the lattice points in star $\bigstar_n$.

The first star ($\bigstar_0=\{{\bf 0}\}$) carries the $\Gamma_1$
irrep.  A star with two lattice points [e.g., $\{(\pm a,0,0)\}$]
carries a 1D irrep of the point group.  A star with four lattice
points [e.g., $\{(\pm a,0,0),(0,\pm b,0)\}$] carries a 2D
representation of the point group.

Where a star $\bigstar_n$ carries a representation that is the direct
sum of irreps carried by other stars, we assume that each lattice
point ${\bf R}_n \in \bigstar_n$ can be decomposed into a sum of
lattice points that transform as the component irreps:
\begin{equation} {\bf R}_n = \frac{1}{2}
\left( n_i {\bf R}_i + n_j {\bf R}_j + n_k {\bf R}_k
\right), \label{eq:reducible_star} \end{equation} where ${\bf R}_i \in
\bigstar_i$, ${\bf R}_j \in \bigstar_j$, and ${\bf R}_k \in
\bigstar_k$, with $\bigstar_i$, $\bigstar_j$, and $\bigstar_k$ being
the lowest-radius stars that carry distinct irreps of the point group
in subspaces of $\spn(\bigstar_n)$ (and we could have
$\bigstar_k=\bigstar_0 \equiv \{{\bf 0}\}$) and $n_i,n_j,n_k \in
\mathbb{Z}$.  The factor of $1/2$ in equation (\ref{eq:reducible_star})
accounts for the possibility that stars that carry reducible
representations may be composed of half-values of lattice points from
irrep stars, as is the case for the body-centred, face-centred, or
base-centred lattices in 3D or the centred rectangular lattice in
2D\@.  Hence the signature of a star $\bigstar_n$
that carries a representation that is the direct sum of irreps carried
by $\bigstar_i$, $\bigstar_j$, and $\bigstar_k$ is that
\begin{equation} (2r_n)^2=n_i^2
r_i^2+n_j^2 r_j^2+n_k^2 r_k^2 \label{eq:test} \end{equation} for some
$n_i,n_j,n_k \in \mathbb{Z}$, which follows from squaring equation
(\ref{eq:reducible_star}), summing over the elements of the point
group, and using the orthogonality theorem for irreps.

The star function $A_n({\bf k})$ cannot be varied independently of the
star functions of the component irrep stars $A_i({\bf k})$, $A_j({\bf
  k})$, and $A_k({\bf k})$.  Hence, before attempting to set a star
function to zero or to minimise it, we must examine its irrep
composition and decide whether it can actually be varied independently
of smaller-radius stars that we have already identified as ``targets''
for zeroing or minimisation.

For example, a star containing eight lattice points $\{(\pm a,0,\pm
c),(0,\pm a,\pm c)\}$ carries a 3D reducible representation, which is
the direct sum of the 1D irrep carried by star $\{(0,0,\pm c)\}$ of
radius $c$ and the 2D irrep carried by star $\{(\pm a,0,0), (0,\pm
a,0)\}$ of radius $a$.  The radius $r$ of star $\{(\pm a,0,\pm
c),(0,\pm a,\pm c)\}$ satisfies $(2r)^2=(2a)^2+(2c)^2$.

In addition, star functions of stars that span independent subspaces
of $\mathbb{R}^3$ [i.e., stars $\bigstar_i$ and $\bigstar_j$ such that
$\spn(\bigstar_i) \cap \spn(\bigstar_j)=\{{\bf 0}\}$] can be varied
independently, even if those stars carry the same representation of
the point group.  A star function in a particular independent subspace
only depends on the component of wavevector in that subspace, and
hence setting the star function to zero or minimising the star
function only provides information about that component of wavevector.
Hence, before looking for star functions that can be varied
independently due to their composition of irreps, we must first look
for stars that together span $\mathbb{R}^3$.

As an example, consider a triclinic lattice.  Each star of lattice
points spans an independent 1D subspace of $\mathbb{R}^3$, and each
subspace carries a 1D irrep of the point group.  We must find the
first (shortest) three stars that, together, span all of 3D space.
This is equivalent to the problem of Minkowski reduction
\cite{Helfrich_1985}, i.e., finding the shortest possible set of
lattice vectors.  We can then determine the component of ${\bf k}$ in
the direction of each of these stars by setting the corresponding star
function to zero (see section \ref{sec:nodes}).

Putting this together, we need to identify at least three ``target''
star functions in order to determine the Baldereschi wavevector.  We
must start by identifying the first (i.e., lowest-radius) stars of
lattice points whose union spans $\mathbb{R}^3$ (i.e., the first
nonzero star of points, then the next star that is not contained in
the space spanned by the first, etc.).  In some cases, such as the
simple cubic lattice, the first nonzero star of lattice points already
spans all of $\mathbb{R}^3$; in other cases (such as the triclinic
lattice or the hexagonal lattice) multiple stars must be found before
their union spans all of $\mathbb{R}^3$.  We can simultaneously set
all these ``target'' star functions to zero through an appropriate
choice of the component of ${\bf k}$ in each of the subspaces spanned
by the target star functions. If the intersections of the spans of the
target stars are just $\{{\bf 0}\}$ then we have stars in independent
subspaces, as is the case for the triclinic lattice discussed above.
If the intersections of the spans of the target stars are finite
dimensional (as is the case for the first two stars of a centred
rectangular lattice with $b/a>\sqrt{3}$) then we have identified stars
composed of different irreps within the vector space that is the union
of the spans of the stars.

Our next task is to look for the shortest stars that can be
independently varied within each subspace.  For each star in
increasing order of radius we test whether the radius $r_n$ satisfies
equation (\ref{eq:test}), where $r_i,\ldots$ are the radii of
already-identified target stars whose span overlaps with the span of
the star that we are testing; if $r_n$ cannot be written in this form
then we have found a new target star.  The stars thus identified
generally carry irreps; the exceptions are situations such as the
centred rectangular lattice in 2D, where the star of lattice points
that includes the centre of the conventional unit cell carries a
reducible representation that is nevertheless chosen to be a target,
because the component irrep-carrying stars have not yet been
identified as targets due to their larger radii.  This is appropriate
because our first priority is to zero or minimise small-radius star
functions, even if they carry reducible representations.  Provided
equation (\ref{eq:test}) is a valid test of whether a star carries an
irrep, this algorithm will find all the irrep-carrying stars in a
subspace together with stars that have shorter radii than the stars
that carry their component irreps.

In those cases where reducible stars have occurred before the stars
carrying their component irreps, we should finally loop over the
targets and search for situations in which equation (\ref{eq:test}) is
satisfied between the radii of all the targets.  Where such relations are
found, we should remove the target star with the highest radius,
leaving us with a set of target stars of minimum radius whose star
functions can be varied independently.  In practice, for 3D Bravais
lattices at least, the targets thus removed are never amongst the
first three targets (which determine the Baldereschi point).

In most cases the target star functions can be set to zero; in other
cases the nodes of the target star functions corresponding to
different irreps within the same subspace do not overlap and hence the
star function for the star with larger radius can only be minimised.
The list of targets we have assembled is ordered such that if a star
can only be minimised, it will be our third target.

Exceptions that may invalidate equation (\ref{eq:test}) occur when
there are accidental degeneracies for particular values of the lattice
parameters (e.g., a 3D hexagonal lattice in which $c=a$).  In practice
this is not a major problem because we are only interested in
low-radius stars and hence there are only a few particular choices of
lattice parameters at which such degeneracies affect the choice of
target stars. More generally it is possible that equation
(\ref{eq:test}) could be accidentally satisfied, but in practice this
does not appear to be a problem in our numerical tests for 3D Bravais
lattices; if target star functions were chosen incorrectly then we
would generally find nonunique solutions for the Baldereschi point,
which are not observed in practice.

Our algorithm for identifying target stars is summarised in appendix
\ref{app:targets}.

\subsection{Nodes of star functions\label{sec:nodes}}

Consider a star $\bigstar$ that carries a 1D irrep.  In this case the
star function is $A({\bf k}) = 2\cos({\bf k}\cdot{\bf R})$, where ${\bf
  R}$ is the sole element of $\bigstar^+$.  The value of the star
function is independent of the component of ${\bf k}$ orthogonal to
${\bf R}$.  The nodes of the star function are the planes ${\bf
  k}\cdot{\bf R}=\pi/2 +l \pi$ for integer $l$.

In a simple orthorhombic or a triclinic lattice, each of the first
three stars that together span $\mathbb{R}^3$ carries a 1D irrep in an
independent subspace, and the nodes of these star functions are planes
that coincide at a single point.  E.g., in an orthorhombic lattice
with lattice vectors $(a,0,0)$, $(0,b,0)$, and $(0,0,c)$, the
Baldereschi point lies at the BZ point with Cartesian coordinates
$\left(\frac{\pi}{2a},\frac{\pi}{2b},\frac{\pi}{2c}\right)$.

Consider a star that carries a 2D representation.  The lattice points
in the star lie in a plane and hence the star function is independent
of the component of ${\bf k}$ in the direction normal to that plane.
So the nodes of the star function are generalised cylinders, i.e.,
surfaces that have translational symmetry in the direction normal to
the plane of the star.

Consider the specific case of a 2D star with fourfold rotational
symmetry.  Without loss of generality, suppose the star is $\{(\pm
a,0,0),(0,\pm a,0)\}$.  So the star function is $A({\bf k})= 2\cos(k_x
a)+2\cos(k_y a)$. The nodes occur when $k_y = \pm (k_x + \pi/a + 2l
\pi/a)$ for integer $l$, i.e., the nodes of the star function are
square prisms in the BZ, with the square cross section oriented the
same way as the square defined by the star of lattice points.  In a
tetragonal lattice with lattice vectors $(a,0,0)$, $(0,a,0)$, and
$(0,0,c)$, the square prism-shaped nodal surfaces of the first two 2D
stars of lattice points coincide at $k_x=\pi/(2a)$ and $k_y=\pi/(2a)$.
The star function due to the first star in the $z$ direction can be
zeroed by setting $k_z=\pi/(2c)$.  Hence the Baldereschi point is at
$\left(\frac{\pi}{2a},\frac{\pi}{2a},\frac{\pi}{2c}\right)$.

If we have two different 2D star functions corresponding to stars in
the same plane then either their nodes coincide on a set of lines
normal to the plane, or their nodes do not coincide.  In the latter
case (e.g., the hexagonal lattice discussed next) we can set one star
function to zero and minimise the second star function; again the
solution is a set of lines normal to the plane.  The star function of
the first star with a component normal to the 2D stars can be set to
zero, determining a point in the BZ\@.

\begin{figure}[htbp]
\centering
\includegraphics[clip,width=8.5cm]{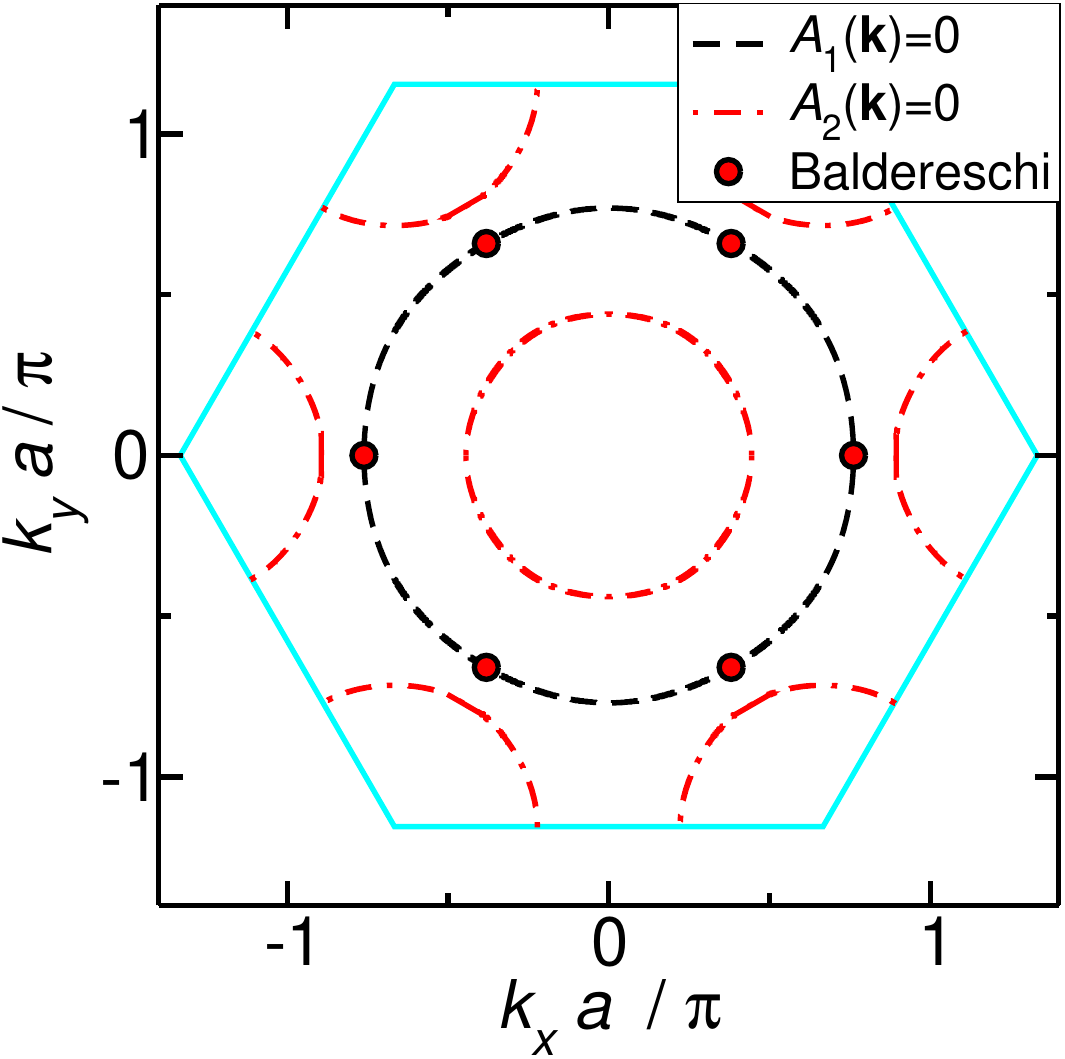}
\caption{\label{fig:hex_nodes} Nodes of the first two nonzero star
  functions $A_1({\bf k})$ and $A_2({\bf k})$ of a hexagonal lattice
  with $c>\sqrt{3}a$ within the first BZ (the cyan hexagon). The
  Baldereschi points are shown by the filled circles.}
\end{figure}

Consider the 3D hexagonal lattice with lattice vectors $(a,0,0)$,
$(-a/2,\sqrt{3}a/2,0)$, and $(0,0,c)$.  For a 2D star with sixfold
rotational symmetry, the nodes of the star function are tubes with
curved cross-sections.  The nodes of the star functions $A_1({\bf
  k})=2\cos(k_x a) + 2\cos(k_x a /2 + \sqrt{3}k_y a/2) + 2\cos(-k_x a
/2 + \sqrt{3}k_y a/2)$ and $A_2({\bf k})=2\cos(3k_x a/2+\sqrt{3}k_y
a/2)+2\cos(\sqrt{3}k_y a) + 2\cos(-3k_x a/2+\sqrt{3}k_y a/2)$ for the
hexagonal lattice with lattice vectors $(a,0,0)$,
$(a/2,\sqrt{3}a/2,0)$, and $(0,0,c)$ are shown in figure
\ref{fig:hex_nodes}, where we have assumed that $c>\sqrt{3}a$, so that
the first two nonzero stars are in-plane.  The nodes have sixfold
rotational symmetry, but one curve is rotated by $30^\circ$ relative
to the other, due to the $30^\circ$ angle between the first two
nonzero stars of lattice points.  The nodes do not cross, so we cannot
set both star functions to zero, but we can minimise the value of
$A_2^2({\bf k})$ on the surface $A_1({\bf k})=0$ to determine the
in-plane component of the Baldereschi point.  The value of $A_2^2({\bf
  k})$ is minimised along the $k_x$-axis.  Hence we can choose $k_y=0$
and solve $A_1=0$ to find $k_x=(2/a) \arccos((\sqrt{3}-1)/2)$.  As
usual, the out-of-plane component $k_z=\pi/(2c)$ is determined by
setting the first star function in the $z$ direction to zero.  The
Baldereschi point reported here, $\left(\frac{2}{a}
\arccos\left(\frac{\sqrt{3}-1}{2}\right),0,\frac{\pi}{2c}\right)$, is
different to the special point reported in reference
\cite{Chadi_1973}; their special point satisfies $A_1({\bf k})=0$, but
maximises rather than minimises $A_2^2({\bf k})$ on the resulting
surface.  However, our result is in agreement with the Baldereschi
point reported in reference \cite{Cunningham_1974}.

We now consider the case of a 2D reducible representation.  In a 2D
centred rectangular lattice with conventional lattice vectors $(a,0)$
and $(0,b)$, the first nonzero star of lattice points is
$\bigstar_1=\{(\pm a/2,\pm b/2)\}$ (four points) when
$1/\sqrt{3}<b/a<\sqrt{3}$.  The corresponding star function is
$A_1({\bf k})=4\cos(k_x a/2)\cos(k_y b/2)$.  The nodes are at
$k_x=\pi/a$ or $k_y=\pi/b$.  If $a<b$ then the second nonzero star of
lattice points is $\bigstar_2=\{(\pm a,0)\}$ and the star function is
$A_2({\bf k})=2\cos(k_x a)$, with nodes at $k_x=\pi/(2a)$.  Hence the
Baldereschi point is ${\bf
  k}_b=\left(\frac{\pi}{2a},\frac{\pi}{b}\right)$.  Likewise, if $a>b$
then ${\bf k}_b=\left(\frac{\pi}{a},\frac{\pi}{2b}\right)$.  If
$b/a>\sqrt{3}$ then $\bigstar_1$ and $\bigstar_2$ are swapped, but the
derivation of the Baldereschi point is otherwise unchanged.  This
construction leads directly to our results for the base-centred and
face-centred orthorhombic and base-centred monoclinic lattices in
table \ref{table:mvps_ctd}.

Finally we consider stars that carry 3D irreps (in practice this only
applies to the cubic lattices originally investigated by Baldereschi
\cite{Baldereschi_1973}).  The nodes of the star functions are a set
of curved surfaces in 3D\@.  If the first nonzero star spans 3D space
then we search for the points of intersection between the nodal
surfaces of the first three target star functions; if there is no such
point of intersection then we look for the minimum of the third target
star function on the surface of intersection of the first and second
target star functions; if there were no such surface of intersection
then we would minimise the second target star function on the nodes of
the first.

\subsection{Method for calculating the Baldereschi mean value point
numerically}

Searching for the intersection of the nodes of target star functions
by solving the three equations $A_{t_1}({\bf k})=A_{t_2}({\bf
  k})=A_{t_3}({\bf k})=0$ with respect to ${\bf k}$ can be performed
numerically by starting from an initial guess at ${\bf k}$ and then
using Newton-Raphson iteration.  If this fails to converge to a
solution then we can search for a solution to $A_{t_1}({\bf
  k})=A_{t_2}({\bf k})=0$ using Newton-Raphson iteration.

Finding a stationary point of the square of target star function
$A_{t_3}({\bf k})$ subject to the constraint $A_{t_1}({\bf
  k})=A_{t_2}({\bf k})=0$ can be performed by introducing Lagrange
multipliers $\lambda_{t_1}$ and $\lambda_{t_2}$ and solving the five
equations $\nabla_{\bf k} \left[ A_{t_3}^2({\bf k})+\lambda_{t_1}
  A_{t_1}({\bf k}) + \lambda_{t_2} A_{t_2}({\bf k}) \right] = {\bf 0}$
and $A_{t_1}({\bf k})=A_{t_2}({\bf k})=0$ with respect to $\{{\bf
  k},\lambda_{t_1},\lambda_{t_2}\}$ using Newton-Raphson iteration.

The method described above searches for constrained stationary points
rather than specifically constrained minima.  To ensure that we find
constrained minima, we start by performing an unconstrained
minimisation using a penalty to enforce approximately the constraints.
Specifically, we minimise $A_{t_3}^2({\bf k})+\left[ A_{t_1}^2({\bf
    k}) + A_{t_2}^2({\bf k}) \right]/\sqrt{\epsilon}$, where
$\epsilon$ is machine precision, using the
Broyden-Fletcher-Goldfarb-Shanno (BFGS) method.  We then use the
Newton-Raphson method to re-impose the constraints exactly, leaving us
very close to the constrained minimum.  Finally, we determine the
constrained stationary point using the method of Lagrange multipliers
described above.

A recent numerical implementation of a method for calculating
Baldereschi points has required the use of tens of millions of random
starting points \cite{Stevanovic_2024}. To sample the BZ efficiently,
we have used a Sobol' quasirandom sequence of ${\bf k}$ values
\cite{Sobol_1976,Bratley_1988}.  The sampled ${\bf k}$ values have
been offset by a constant vector, to ensure that the initial ${\bf k}$
are not symmetry-equivalent.  The loop over initial ${\bf k}$ can
performed in parallel.  However, in practice, nonglobal minima of $g$
are only found in the case of the rhombohedral lattice; and even then,
only a handful of initial wavevectors are required to find the global
minimum.
In fact, once the target stars have been identified, it is often
possible to solve for the Baldereschi point analytically using simple
geometrical arguments.  Nevertheless, we have found it extremely
useful to be able to check the analytical results given in tables
\ref{table:mvps} and \ref{table:mvps_ctd} numerically.  A small
computer program was written to find Baldereschi points numerically,
and is included in the \textsc{casino} distribution
\cite{Needs_2020,CASINO}.  The algorithm is summarised in appendix
\ref{app:numerics}.

\section{Baldereschi mean value point for each Bravais lattice}

In table \ref{table:mvps} we report the Baldereschi mean-value points
for the Bravais lattices in the cubic, hexagonal, and tetragonal
families.

\begin{table*}
\centering
\caption{Baldereschi mean-value points ${\bf k}_{\rm b}$ for the
  cubic, hexagonal, and tetragonal families of Bravais lattices.
  Baldereschi points for the orthorhombic, monoclinic, and triclinic
  families of Bravais lattices can be found in table
  \ref{table:mvps_ctd}. ${\bf b}_1=2\pi {\bf a}_2 \times {\bf
    a}_3/[{\bf a}_1 \cdot ({\bf a}_2 \times {\bf a}_3)]$, ${\bf
    b}_2=2\pi {\bf a}_3 \times {\bf a}_1/[{\bf a}_1 \cdot ({\bf a}_2
    \times {\bf a}_3)]$, and ${\bf b}_3=2\pi {\bf a}_1 \times {\bf
    a}_2/[{\bf a}_1 \cdot ({\bf a}_2 \times {\bf a}_3)]$ are primitive
  reciprocal lattice vectors, where ${\bf a}_1$, ${\bf a}_2$, and
  ${\bf a}_3$ are primitive real lattice vectors. $a$, $b$, $c$,
  etc.\ are lattice parameters.  We define the constants
  $\phi=\frac{1}{2\pi} \arccos\left( \frac{\sqrt{3}-1}{2} \right)$ and
  $\alpha_{\rm d}=\arccos(-1/8)$.  In all cases star functions
  $A_{t_1}({\bf k}_{\rm b})$ and $A_{t_2}({\bf k}_{\rm b})$ are set to
  zero.\label{table:mvps}}
\begin{tabular}{lccc}
\hline

Lattice & \parbox{2.5cm}{\raggedright Lattice vectors ${\bf a}_1$,
  ${\bf a}_2$, and ${\bf a}_3$} & Baldereschi point ${\bf k}_{\rm b}$
& $A_{t_3}({\bf k}_{\rm b})$ \\

\hline

Simple cubic & \parbox{2cm}{$(a,0,0)$ \\ $(0,a,0)$ \\ $(0,0,a)$} &
$\frac{1}{4}{\bf b}_1 + \frac{1}{4}{\bf b}_2 + \frac{1}{4}{\bf b}_3$ &
$0$ \\

\hline

\parbox{2cm}{\raggedright Face-centred cubic}
& \parbox{2cm}{$\left(0,\frac{a}{2},\frac{a}{2}\right)$
  \\ $\left(\frac{a}{2},0,\frac{a}{2}\right)$
  \\ $\left(\frac{a}{2},\frac{a}{2},0\right)$} & $\begin{array}{c}
  0.1476669075533311 \, {\bf b}_1 \\ {} + 0.3111505578912695 \, {\bf
    b}_2 \\ {} + 0.4588174654446007 \, {\bf b}_3 \end{array}$ &
$4.404$ \\

\hline

\parbox{2cm}{\raggedright Body-centred cubic}
& \parbox{2cm}{$\left(-\frac{a}{2},\frac{a}{2},\frac{a}{2}\right)$
  \\ $\left(\frac{a}{2},-\frac{a}{2},\frac{a}{2}\right)$
  \\ $\left(\frac{a}{2},\frac{a}{2},-\frac{a}{2}\right)$} &
$\frac{1}{4} {\bf b}_1 + \frac{1}{4} {\bf b}_2 + \frac{7}{12} {\bf
  b}_3$ & $-3$ \\

\hline

\parbox{2cm}{Hexagonal} & \parbox{2cm}{$(a,0,0)$
  \\ $\left(-\frac{a}{2},\frac{\sqrt{3}a}{2},0\right)$ \\ $(0,0,c)$} &
$\phi {\bf b}_1 + \phi {\bf b}_2 + \frac{1}{4} {\bf b}_3$ & $-1.608$
\\

\hline

\parbox{2cm}{\raggedright Hexagonal (alternative)} & \parbox{2cm}{$(a,0,0)$
  \\ $\left(\frac{a}{2},\frac{\sqrt{3}a}{2},0\right)$ \\ $(0,0,c)$} &
$2\phi {\bf b}_1 + \phi {\bf b}_2 + \frac{1}{4} {\bf b}_3$ & $-1.608$
\\

\hline

\parbox{2cm}{\raggedright Rhombohedral (obverse, hexagonal cell)} &
\parbox{2cm}{$(a,0,0)$ \\ $\left(-\frac{a}{2},\frac{\sqrt{3}a}{2},0
  \right)$ \\ $\left( \frac{a}{2}, \frac{\sqrt{3}a}{6},\frac{c}{3}
  \right)$} & $\left\{ \begin{array}{ll} \phi {\bf b}_1 + \phi {\bf
    b}_2 + \left( \frac{1}{4} + \phi \right) {\bf b}_3 & \mbox{if~}
  c>\sqrt{6}a \\[1ex] \frac{1}{3}{\bf b}_2 + \frac{1}{3} {\bf b}_3 &
  \mbox{if~} \sqrt{\frac{3}{2}}a<c<\sqrt{6}a \\[1ex] \left( 1 -
  3\phi \right) {\bf b}_2 + \left( \frac{1}{2} -\phi \right) {\bf b}_3 &
  \mbox{if~} a < c < \sqrt{\frac{3}{2}}a \\[1ex] \frac{1}{3} {\bf b}_1
  + \frac{1}{3} {\bf b}_2 + \frac{5}{12} {\bf b}_3 & \mbox{if~}
  c<a \end{array} \right.$ & $\left\{\begin{array}{l} -1.608 \\[1em]
-3 \\[1em] -1.608 \\[1em] 0 \end{array} \right.$ \\

\hline

\parbox{2cm}{\raggedright Rhombohedral (reverse, hexagonal cell)} &
\parbox{2cm}{$(a,0,0)$ \\ $\left(-\frac{a}{2},\frac{\sqrt{3}a}{2},0
  \right)$ \\ $\left( 0, \frac{a}{\sqrt{3}},\frac{c}{3} \right)$} &
$\left\{ \begin{array}{ll} \phi {\bf b}_1 + \phi {\bf b}_2 + \left(
  \frac{1}{4} + \phi \right){\bf b}_3 & \mbox{if~} c>\sqrt{6}a \\[1ex]
  \frac{1}{3}{\bf b}_1 + \frac{1}{3} {\bf b}_3 & \mbox{if~}
  \sqrt{\frac{3}{2}}a<c<\sqrt{6}a \\[1ex] \left( 3\phi -1 \right)
  {\bf b}_1 + \phi {\bf b}_3 & \mbox{if~} a < c <
  \sqrt{\frac{3}{2}}a \\[1ex] \frac{1}{3} {\bf b}_1 + \frac{1}{3} {\bf
    b}_2 + \frac{1}{4} {\bf b}_3 & \mbox{if~} c<a \end{array} \right.$
& $\left\{\begin{array}{l} -1.608 \\[1em] -3 \\[1em] -1.608 \\[1em]
0 \end{array} \right.$ \\

\hline

\parbox{2cm}{\raggedright Rhombohedral (rhombohedral cell)}
& \parbox{4cm}{$(a,0,0)$
  \\ $\left(a\cos(\alpha),a\sin(\alpha),0\right)$
  \\ $\left( \begin{array}{c} a\cos(\alpha), \\ a
    \frac{\cos(\alpha)[1-\cos(\alpha)]}{\sin(\alpha)}, \\ a
    \frac{\sqrt{1-\cos^2(\alpha)\left[ 3 - 2\cos(\alpha)
          \right]}}{\sin(\alpha)} \end{array} \right)$} &
$\left\{ \begin{array}{ll} \left(\frac{1}{4} - \phi \right) {\bf b}_1
  + \frac{1}{4} {\bf b}_2 + \left( \frac{1}{4}+\phi \right) {\bf b}_3
  & \mbox{if~} 0< \alpha< \frac{\pi}{3} \\[1ex] \frac{1}{3} {\bf b}_2
  + \frac{1}{3} {\bf b}_3 & \mbox{if~} \frac{\pi}{3} < \alpha <
  \frac{\pi}{2} \\[1ex] \left( 1 - 2\phi \right) {\bf b}_1 + \phi {\bf
    b}_2 + \phi {\bf b}_3 & \mbox{if~} \frac{\pi}{2} < \alpha <
  \alpha_{\rm d} \\[1ex] \frac{1}{12} {\bf b}_1 +
  \frac{5}{12} {\bf b}_2 + \frac{3}{4} {\bf b}_3 &
  \mbox{if~}\alpha_{\rm d} < \alpha <
  \frac{2\pi}{3} \end{array} \right.$ & $\left\{ \begin{array}{l}
  -1.608 \\[1em] -3 \\[1em] -1.608 \\[1em] 0 \end{array} \right.$ \\

\hline

\parbox{2cm}{\raggedright Simple tetragonal} & \parbox{2cm}{$(a,0,0)$
  \\ $(0,a,0)$ \\ $(0,0,c)$} & $\frac{1}{4} {\bf b}_1 + \frac{1}{4}
       {\bf b}_2 + \frac{1}{4} {\bf b}_3$ & $0$ \\

\hline

\parbox{2cm}{\raggedright Body-centred tetragonal} &
\parbox{2cm}{$\left(-\frac{a}{2},\frac{a}{2},\frac{c}{2}\right)$
  \\ $\left(\frac{a}{2},-\frac{a}{2},\frac{c}{2}\right)$
  \\ $\left(\frac{a}{2},\frac{a}{2},-\frac{c}{2}\right)$} &
$\left\{ \begin{array}{ll} \frac{1}{4}{\bf b}_1 + \frac{1}{4} {\bf
    b}_2 & \mbox{if~} c>\sqrt{2}a \\[1ex] \frac{1}{8} {\bf b}_1 +
  \frac{5}{8} {\bf b}_2 + \frac{3}{8} {\bf b}_3 & \mbox{if~} c<\sqrt{2}a \end{array} \right.$ & $0$ \\

\hline
\end{tabular}
\end{table*}

\begin{table*}
\centering
\caption{As table \ref{table:mvps}, but for the orthorhombic,
  monoclinic, and triclinic families of Bravais lattices.
The lattice vectors for the triclinic lattice are assumed to be
Minkowski-reduced, i.e.\ the shortest possible lattice vectors have
been chosen, while the lattice vectors for the monoclinic and
base-centred monoclinic lattices have been chosen to have the shortest
possible value of $|{\bf a}_3|$.\label{table:mvps_ctd}}
\begin{tabular}{lccc}
\hline

Lattice & \parbox{2.5cm}{\raggedright  Lattice vectors \\ ${\bf a}_1$,
  ${\bf a}_2$, and ${\bf a}_3$} & Baldereschi point ${\bf k}_{\rm b}$
& $A_{t_3}({\bf k})$ \\

\hline

\parbox{3cm}{\raggedright Simple orthorhombic} & \parbox{2cm}{$(a,0,0)$
  \\ $(0,b,0)$ \\ $(0,0,c)$} & $\frac{1}{4}{\bf b}_1 + \frac{1}{4}
       {\bf b}_2 + \frac{1}{4} {\bf b}_3$ & $0$ \\

\hline

\parbox{3cm}{\raggedright Base-centred orthorhombic}
& \parbox{2cm}{$\left(\frac{a}{2},\frac{b}{2},0\right)$
  \\ $\left(-\frac{a}{2},\frac{b}{2},0\right)$ \\ $(0,0,c)$} &
$\left\{ \begin{array}{ll} \frac{1}{8} {\bf b}_1 + \frac{3}{8} {\bf b}_2 +
  \frac{1}{4} {\bf b}_3 & \mbox{if~} b>a \\[1ex] \frac{1}{8}{\bf
    b}_1 + \frac{5}{8} {\bf b}_2 + \frac{1}{4} {\bf b}_3 & \mbox{if~}
  b<a \end{array} \right.$ & $0$ \\

\hline

\parbox{3cm}{\raggedright Body-centred orthorhombic}
& \parbox{2cm}{$\left(-\frac{a}{2},\frac{b}{2},\frac{c}{2} \right)$
  \\ $\left(\frac{a}{2},-\frac{b}{2},\frac{c}{2} \right)$
  \\ $\left(\frac{a}{2},\frac{b}{2},-\frac{c}{2} \right)$} &
$\left\{ \begin{array}{ll} \frac{1}{4}{\bf b}_2 + \frac{1}{4}{\bf b}_3 &
  \mbox{if~}b,c<a \\[1ex] \frac{1}{4}{\bf b}_1 + \frac{1}{4}{\bf b}_3 &
  \mbox{if~}a,c<b \\[1ex] \frac{1}{4}{\bf b}_1 + \frac{1}{4} {\bf b}_2 &
  \mbox{if~}a,b<c \end{array} \right.$ & $0$ \\

\hline

\parbox{3cm}{\raggedright Face-centred orthorhombic}
& \parbox{2cm}{$\left(0,\frac{b}{2},\frac{c}{2} \right)$
  \\ $\left(\frac{a}{2},0,\frac{c}{2} \right)$
  \\ $\left(\frac{a}{2},\frac{b}{2},0 \right)$} &
$\left\{ \begin{array}{ll} \frac{1}{2} {\bf b}_1 +\frac{1}{8}{\bf b}_2
  + \frac{1}{8} {\bf b}_3 & \mbox{if~}a<b,c \\[1ex] \frac{1}{8} {\bf
    b}_1 +\frac{1}{2}{\bf b}_2 + \frac{1}{8} {\bf b}_3 &
  \mbox{if~}b<a,c \\[1ex] \frac{1}{8} {\bf b}_1 +\frac{1}{8}{\bf b}_2
  + \frac{1}{2} {\bf b}_3 & \mbox{if~}c<a,b \end{array} \right.$ & 0
\\

\hline

\parbox{3cm}{\raggedright Monoclinic (minimum possible $|{\bf a}_3|$)}
& \parbox{2cm}{$(a,0,0)$ \\ $(0,b,0)$ \\ $(c_1,0,c_2)$} &
$\frac{1}{4}{\bf b}_1 + \frac{1}{4} {\bf b}_2 + \frac{1}{4}{\bf b}_3$
& $0$ \\

\hline

\parbox{3cm}{\raggedright Base-centred monoclinic (minimum possible
  $|{\bf a}_3|$)}
& \parbox{2cm}{$\left(\frac{a}{2},\frac{b}{2},0\right)$
  \\ $\left(-\frac{a}{2},\frac{b}{2},0\right)$ \\ $(c_1,0,c_2)$} &
$\left\{ \begin{array}{ll} \frac{1}{8} {\bf b}_1 + \frac{3}{8} {\bf
    b}_2 + \frac{1}{4} {\bf b}_3 & \mbox{if~} b>a \\[1ex] \frac{1}{8}
  {\bf b}_1 + \frac{5}{8} {\bf b}_2 + \frac{1}{4} {\bf b}_3 &
  \mbox{if~} b<a \end{array} \right.$ & $0$ \\

\hline

\parbox{3cm}{\raggedright Triclinic (Minkowski-reduced)}
& \parbox{2cm}{$(a,0,0)$ \\ $(b_1,b_2,0)$ \\ $(c_1,c_2,c_3)$} &
$\frac{1}{4}{\bf b}_1 + \frac{1}{4} {\bf b}_2 + \frac{1}{4} {\bf b}_3$
& $0$ \\

\hline
\end{tabular}
\end{table*}

The simple cubic, body-centred cubic, and face-centred cubic lattices
were studied in Baldereschi's original work \cite{Baldereschi_1973}.
Our results are in agreement with his results, although we report the
face-centred cubic mean-value point with very much greater precision.

The Baldereschi mean-value points of the five 2D Bravais lattices were
evaluated in reference \cite{Cunningham_1974}, and the extension of
these results to the 3D hexagonal and simple tetragonal lattices
\cite{Bashenov_1977,Chulkov_1979,Evarestov_1983}, as well as the
simple orthorhombic, base-centred orthorhombic, monoclinic, and
base-centred monoclinic lattices is straightforward.

In the rhombohedral lattice we get crossovers in behaviour (i) when
$|{\bf a}_3|=|{\bf a}_1|$, i.e., when $c/a=\sqrt{6}$; (ii) when
$|2{\bf a}_3|=|{\bf a}_1|$, i.e., when $c/a=\sqrt{3/2}$; and (iii)
when $|3{\bf a}_3-{\bf a}_1|=|{\bf a}_1|$, i.e., when $c/a=1$.  The
internal angle $\alpha$ of the rhombohedral primitive cell is the
angle between lattice vectors ${\bf a}_3$ and ${\bf a}_3-{\bf a}_2$;
hence it is easy to relate these crossovers to values of the
rhombohedral angle $\alpha$.  Results for the rhombohedral lattice in
the rhombohedral unit cell are presented in reference
\cite{Bashenov_1977}, but disagree with our results in the
intermediate regions ($60^\circ < \alpha< 107.5^\circ$), presumably
due to the selection of target star functions.  Our results for the
body-centred tetragonal lattice are in agreement with the results
presented in reference \cite{Bashenov_1977}.

The Baldereschi mean value points of the orthorhombic, monoclinic, and
triclinic families of Bravais lattices are shown in table
\ref{table:mvps_ctd}.  We are not aware of previously reported
Baldereschi mean value points for these lattices, other than the
numerical results for some particular cases given in reference
\cite{Stevanovic_2024}.  However, in general the mean value points
follow straightforwardly from the corresponding 2D and 1D results.

The numerical results of reference \cite{Stevanovic_2024} do not
generally agree with our results.  A clear example of this is provided
by the orthorhombic lattice.  In this case the analytical result for
the Baldereschi point ${\bf k}_{\rm b}=\frac{1}{4}{\bf
  b}_1+\frac{1}{4}{\bf b}_2+\frac{1}{4}{\bf b}_3$ follows immediately
from the independence of the three Cartesian directions; however, two
different numerical values are reported in reference
\cite{Stevanovic_2024}, both of which are in disagreement with the
analytical result. The likely reason for the disagreement is that the
algorithm of reference \cite{Stevanovic_2024} does not start by
identifying the first three stars whose union spans $\mathbb{R}^3$.

\section{Three examples examining the effectiveness of the Baldereschi
point\label{sec:examples}}

\subsection{Hexagonal boron nitride}

Hexagonal boron nitride (hBN) is an insulator that consists of
honeycomb-structured hBN monolayers vertically stacked in an AA$'$
sequence.  The Bravais lattice is hexagonal, with the lattice
constants being $a=2.504$ {\AA} and $c=6.6612$ {\AA}
\cite{Lynch_1966}.  To investigate the performance of the Baldereschi
point as it would be used in QMC calculations, we examine plane-wave
DFT energies calculated using \textsc{castep} \cite{Clark_2005} with
the PBE exchange-correlation functional, ultrasoft pseudopotentials, a
plane-wave cutoff energy of 28.8115 Ha, and $n \times n \times n$
grids of ${\bf k}$ points offset from the Brillouin zone centre
$\Gamma$ by the twist ${\bf k}_{\rm s}$.  In a QMC calculation, the $n
\times n \times n$ ${\bf k}$-point grid would be unfolded into an $n
\times n \times n$ supercell with a single Bloch wavevector ${\bf
  k}_{\rm s}$.  Exploring momentum quantisation effects using DFT is
convenient because (i) there are no random errors in the energy at
each twist and (ii) momentum quantisation (i.e., ${\bf k}$-point
sampling errors) are the only source of finite-size effects.

\begin{figure}[htbp]
\centering
\includegraphics[clip,width=8.5cm]{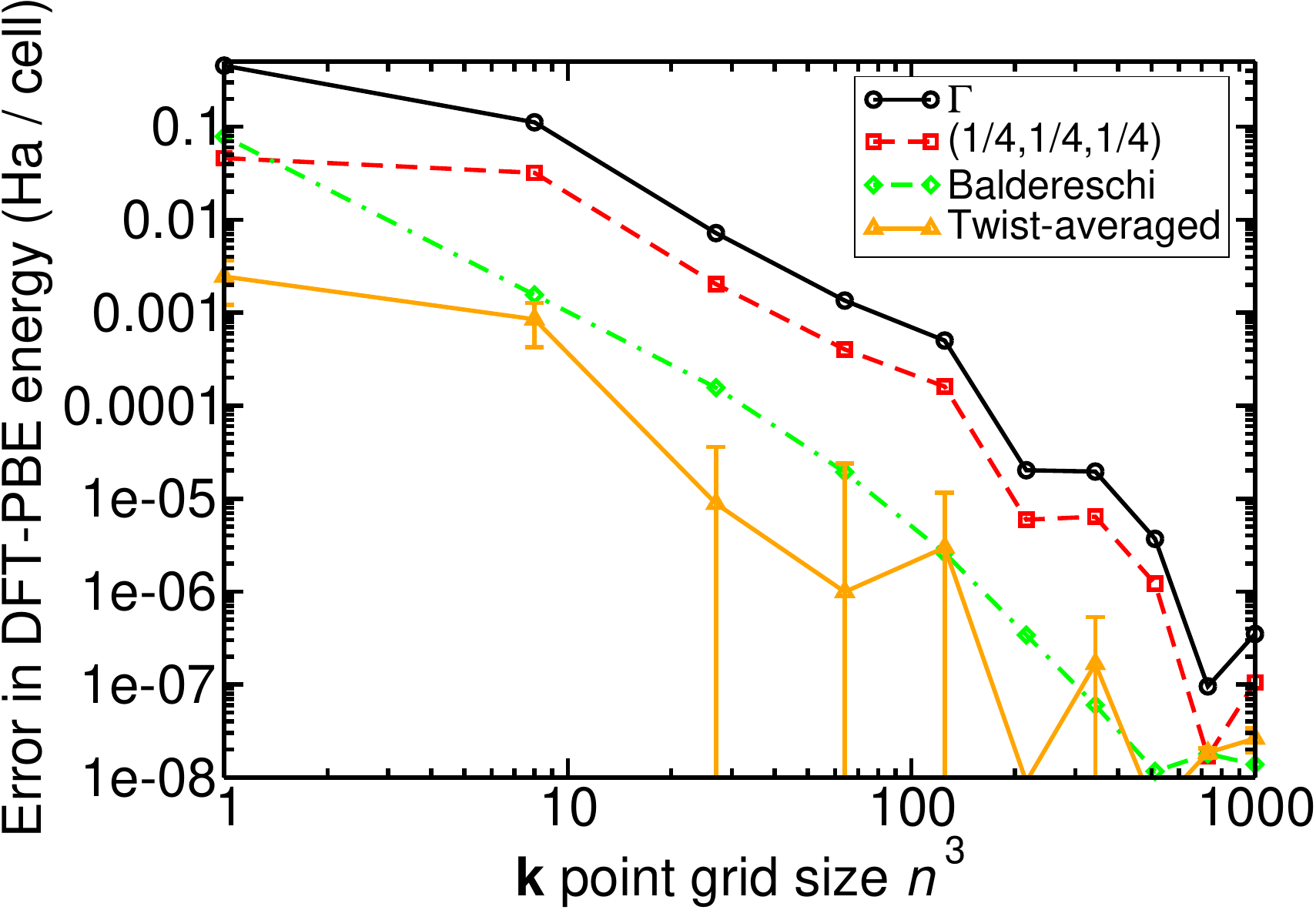}
\caption{\label{fig:hBN_Eerr_v_nk} Difference between the DFT-PBE
  energy of bulk hBN with a $n \times n \times n$ ${\bf k}$-point grid
  and the DFT-PBE energy with a large $123 \times 123 \times 40$ ${\bf
    k}$-point grid. Results are shown for different twists ${\bf
    k}_{\rm s}$ (i.e., offsets to the grid of ${\bf k}$ points).  The
  labels ``$\Gamma$'', ``$(1/4,1/4,1/4)$'', and ``Baldereschi''
  indicate the position of ${\bf k}_{\rm s}$ within the Brillouin zone
  of the unfolded $n \times n \times n$ supercell.  The twist-averaged
  results are averaged over randomly sampled ${\bf k}_{\rm s}$ in the
  supercell Brillouin zone; the error bars indicate the standard error
  in the twist-averaged energy.}
\end{figure}

Figure \ref{fig:hBN_Eerr_v_nk} shows that twist averaging is very
effective at removing momentum quantisation errors in insulators.
Even with a $3 \times 3 \times 3$ ${\bf k}$-point grid, the
twist-averaged energy per primitive cell is already within about 10
$\mu$Ha of the fully converged result (c.f., ``chemical accuracy'' is
about 1.6 mHa).  For all but the $1 \times 1 \times 1$ grid, choosing
${\bf k}_{\rm s}$ to lie at the Baldereschi point of the supercell
Brillouin zone gives DFT energies that are only slightly less accurate
than the twist-averaged results.  Furthermore, the momentum
quantisation error at the Baldereschi point falls off relatively
smoothly at a rate faster than $N^{-2}$, where $N=n^3$ is the size of
the unfolded supercell.  As the supercell size increases, the
supercell Brillouin zone becomes smaller and hence the dependence of
the results on the twist becomes weaker; we therefore expect the
Baldereschi point to be increasingly effective in large supercells,
facilitating extrapolation to infinite supercell size.
\textsc{Castep}'s standard ${\bf k}$-point grids alternate between the
twist ${\bf k}_{\rm s}$ lying at $\Gamma$ (for odd $n$) and at the
point with fractional coordinates $(1/4,1/4,1/4)$ in the supercell
Brillouin zone (for even $n$).  For a given size of grid, choosing the
twist to be the supercell Baldereschi point makes the energy more than
an order of magnitude closer to the converged result.

\subsection{Monolayer graphene}

Monolayer graphene is a 2D allotrope of carbon with a honeycomb
structure.  It has a 2D hexagonal Bravais lattice with lattice
parameter $a=2.46$ {\AA}\@.  Unlike hBN, it is a semimetal with zero
band gap.  Here we compare twist-averaged QMC with non-twist-averaged
QMC at the 2D Baldereschi point for (i) the total energy and (ii) the
out-of-plane polarisability, as obtained by placing monolayer graphene
in an out-of-plane external electric field of strength 0.002 Ha/bohr/e
and calculating the resulting dipole moment.

The calculations are performed in a supercell of $3 \times 3$
primitive cells subject to twisted 2D-periodic boundary conditions.
The carbon atoms are represented by Trail-Needs Dirac-Fock
pseudopotentials \cite{Trail_2005a,Trail_2005b}.  We use a
Slater-Jastrow trial wave function in which the Slater determinants
contain PBE orbitals generated using \textsc{castep} and
re-represented in a blip (B-spline) basis set, and the Jastrow factor
contains polynomial electron-electron, electron-nucleus and
electron-electron-nucleus terms \cite{Drummond_2004} that are
optimised by energy minimisation \cite{Umrigar_2007}.  Diffusion Monte
Carlo (DMC) simulations are performed at time steps of 0.01 and 0.04
Ha$^{-1}$ and the results are extrapolated linearly to zero time step,
with the target population being varied in inverse proportion to the
time step.  In the twist-averaged calculation we use 24 random twists,
with the Jastrow factor being reoptimised at each twist.  The
polarisability is evaluated by extrapolated estimation (twice the DMC
mixed expectation value minus the variational Monte Carlo expectation
value) to eliminate errors that are first order in the error in the
trial wave function.  Our QMC calculations are performed with
\textsc{casino} \cite{Needs_2020}.

Our results for the energy and polarisability of graphene are shown in
Table \ref{table:graphene_results}.  The standard deviation of the DMC
energies at the 24 random twists is 6.0 mHa per cell, which is large
compared with the error bars on the DMC energy at each twist (about 25
$\mu$Ha per cell); hence the standard deviation quantifies the spread
of DMC energy with twist.  The difference between the energy at the
Baldereschi point and the na\"{\i}ve twist averaged energy is
$-0.9(12)$ mHa per cell, which is statistically insignificant, and
much less than the spread of energy as a function of twist.  Likewise
the difference between the polarisability at the Baldereschi point and
the twist-averaged polarisability is $-0.02(10)$ bohr$^3$ per cell,
which is again statistically insignificant and smaller than the
standard deviation $0.33$ bohr$^3$ per cell of the polarisability as a
function of twist.

\begin{table}
\centering
\caption{DMC mixed estimate of the ground-state energy and QMC
  extrapolated estimate of the out-of-plane polarisability of a $3
  \times 3$ supercell of monolayer graphene.  Both na\"{\i}ve and
  corrected twist-averaged results are presented; the former are a
  simple average of the QMC results obtained at different twists,
  while the latter are evaluated using equation \ref{eq:ta_fit}. The
  DMC energy at the Baldereschi point corrected by the difference
  between the DFT-PBE energy with a fine ${\bf k}$-point grid and the
  DFT-PBE energy with a $3 \times 3$ ${\bf k}$-point grid offset by
  the supercell Baldereschi point is also
  given.\label{table:graphene_results}}
\begin{tabular}{lcc}
\hline

Twist & DMC energy (Ha/cell) & QMC polarisability (bohr$^3$/cell) \\

\hline

Baldereschi &   $-11.367321(5)$ & $5.66(8)$ \\

Twist average (na\"{\i}ve) & $-11.3664(12)$ & $5.68(7)$ \\

\hline

Baldereschi (corrected) & $-11.366436(5)$ & \\

Twist average (corrected) & $-11.36666(3)$ & \\

\hline
\end{tabular}
\end{table}

To obtain a more precise and accurate twist-averaged energy, we use
the corresponding DFT energies as a control variate.  We fit
\begin{equation} E({\bf k}_{\rm s}) = \bar{E} + b\left[
E_{\rm DFT}({\bf k}_{\rm s}) - E_{\rm DFT}(\infty)
\right] \label{eq:ta_fit} \end{equation} to the QMC energies $E({\bf
  k}_{\rm s})$ as a function of twist ${\bf k}_{\rm s}$, where the
corrected twist-averaged energy $\bar{E}$ and $b$ are fitting
parameters and $E_{\rm DFT}({\bf k}_{\rm s})$ is the DFT-PBE energy
with a $3 \times 3$ ${\bf k}$-point grid offset by ${\bf k}_{\rm s}$
and $E_{\rm DFT}(\infty)$ is the DFT energy with a fine 2D ${\bf
  k}$-point grid (in this case a $101 \times 101$ mesh).  This
approach improves accuracy as well as precision, because it implicitly
corrects for the difference between the twist-averaged energy and the
energy without momentum quantisation errors, due to our use of $E_{\rm
  DFT}(\infty)$ rather than the twist-averaged DFT energy in equation
(\ref{eq:ta_fit}).

To make a fair comparison, we also correct the DMC energy at the
supercell Baldereschi point ${\bf k}_{\rm b}$ by adding $E_{\rm
  DFT}(\infty) - E_{\rm DFT}({\bf k}_{\rm b})$.  The difference
between the corrected twist-averaged DMC energy and the corrected
Baldereschi-point DMC energy (both shown in table
\ref{table:graphene_results}) is $0.22(3)$ mHa per cell, more than an
order of magnitude smaller than the standard deviation of the DMC
energy as a function of twist.

In summary, the DMC total energy and the QMC out-of-plane
polarisability of a $3 \times 3$ supercell of graphene at the
supercell Baldereschi point are both far closer to the corresponding
twist-averaged results than is the case for a typical randomly chosen
twist.  The effectiveness of the Baldereschi point increases at larger
supercell sizes, where the dependence on the twist becomes weaker;
hence the Baldereschi point could be used instead of twist averaging
when extrapolating QMC results for graphene to infinite supercell
size.

\subsection{3D free electron gas}

The free electron gas provides an excellent test case for exploring
momentum quantisation finite-size errors in metallic systems, because
calculations are simple and analytical results are available in the
thermodynamic limit.  Here we examine a two-component free electron
gas in a simple cubic simulation cell subject to twisted periodic
boundary conditions with twist ${\bf k}_{\rm s}$.

The total (kinetic) energy per electron of the free electron gas in an
$N$-electron cell (which is equal to the Hartree-Fock kinetic energy
per electron of an interacting homogeneous electron gas) is
\begin{equation} T(N) = \frac{2}{N} \sum_{\mbox{occupied~}{\bf k}}
\frac{\hbar^2 k^2}{2m_{\rm e}}, \end{equation} where the allowed
wavevectors are of the form ${\bf k}={\bf G}+{\bf k}_{\rm s}$, where
the $\{{\bf G}\}$ are the simulation-cell reciprocal lattice points,
i.e., the allowed wavevectors lie on the grid of simulation-cell
reciprocal lattice points offset by the twist ${\bf k}_{\rm s}$.  The
ground-state occupied wave vectors are the $N/2$ wavevectors ${\bf k}$
closest to the origin of reciprocal space.  As the thermodynamic limit
$N \to \infty$ is approached at fixed density, the grid of wavevectors
becomes increasingly fine and, as is well known, the kinetic energy
per electron tends to
\begin{equation} T(\infty) = \frac{2}{n} \int_0^{k_{\rm F}} \frac{1}{(2\pi)^3}
\frac{\hbar^2 k^2}{2m_{\rm e}} 4\pi k^2 \, dk = \frac{3\hbar^2 k_{\rm
    F}^2}{10m_{\rm e}}, \end{equation} where $k_{\rm F}=(3\pi^2
n)^{1/3}$ is the Fermi wavevector and $n$ is the number density.

\begin{figure}[htbp]
\centering
\includegraphics[clip,width=8.5cm]{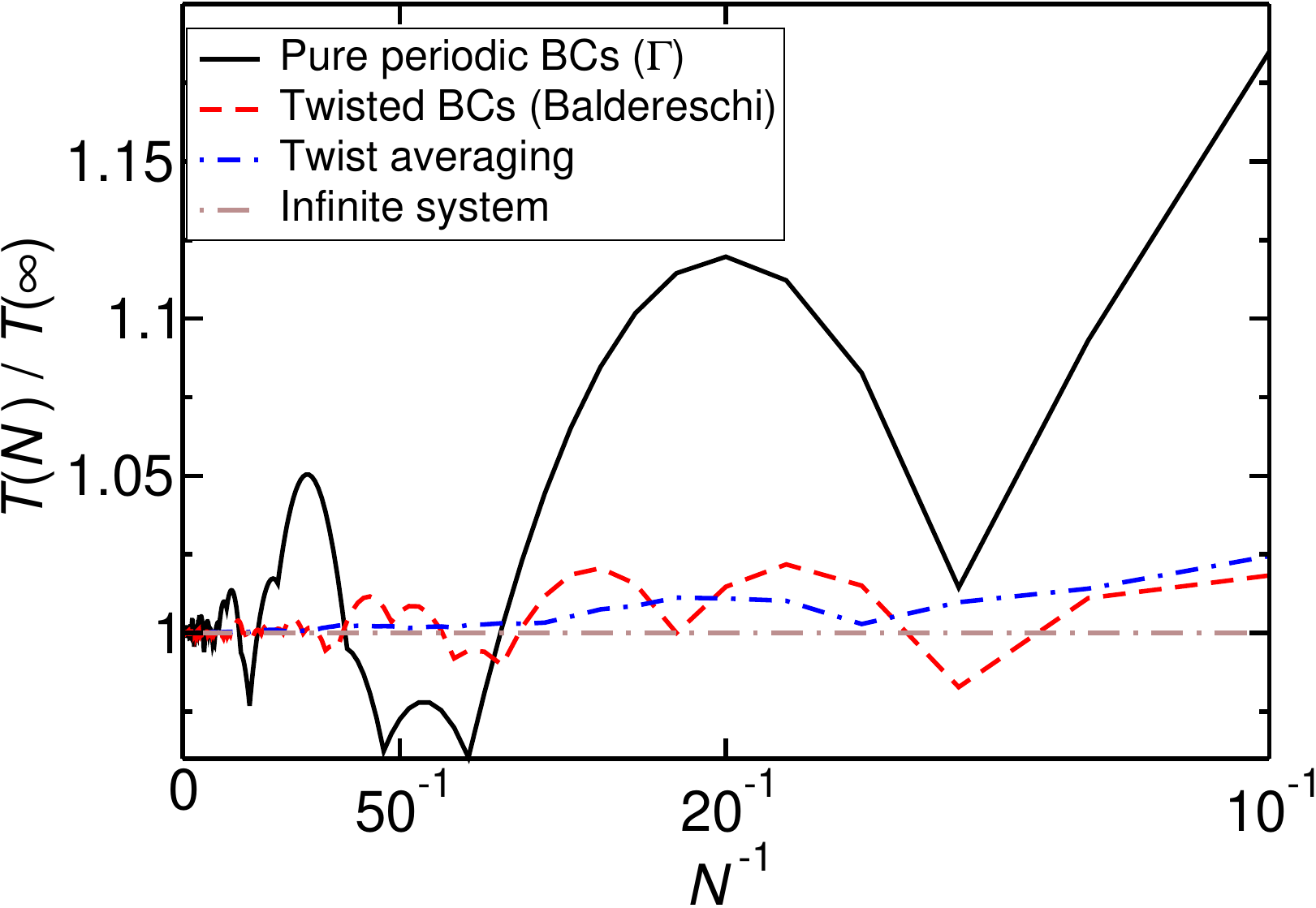}
\caption{\label{fig:HEG3D_SC_para_HF_T_v_N} Kinetic energy per
  electron $T(N)$ of a two-component free electron gas in a simple
  cubic simulation cell against the reciprocal of the system size $N$.
  Results are shown for finite simulation cells subject to pure
  periodic boundary conditions, twisted periodic boundary condition
  with the twist at the simulation cell Baldereschi point, and
  twist-averaging.}
\end{figure}

In figure \ref{fig:HEG3D_SC_para_HF_T_v_N} we plot the free electron
gas kinetic energy per electron against inverse system size.  The
kinetic energy in a finite cell oscillates in a quasirandom manner
before eventually tending to the thermodynamic limit.  Twist-averaging
converts the sum over discrete ${\bf k}$ into an integral over a
polyhedron with the same volume as the Fermi sphere; it therefore
greatly suppresses the fluctuations with system size and gives a
small, positive finite-size error, since the Fermi sphere is by
definition the shape of occupied region that minimises the kinetic
energy per electron.

The kinetic energies with pure periodic boundary conditions and with
twisted periodic boundary conditions at the Baldereschi point
oscillate quasirandomly about the twist-averaged kinetic energy due to
shell-filling effects, as stars of symmetry-equivalent allowed
wavevectors are progressively occupied.  However, it is clear that the
oscillations are an order of magnitude smaller at the Baldereschi
point.  Hence the kinetic energy per electron in a finite system at
the Baldereschi point is in general very much closer to the
infinite-system value than is the kinetic energy per electron with
pure periodic boundary conditions.

\begin{figure}[htbp]
\centering
\includegraphics[clip,width=8.5cm]{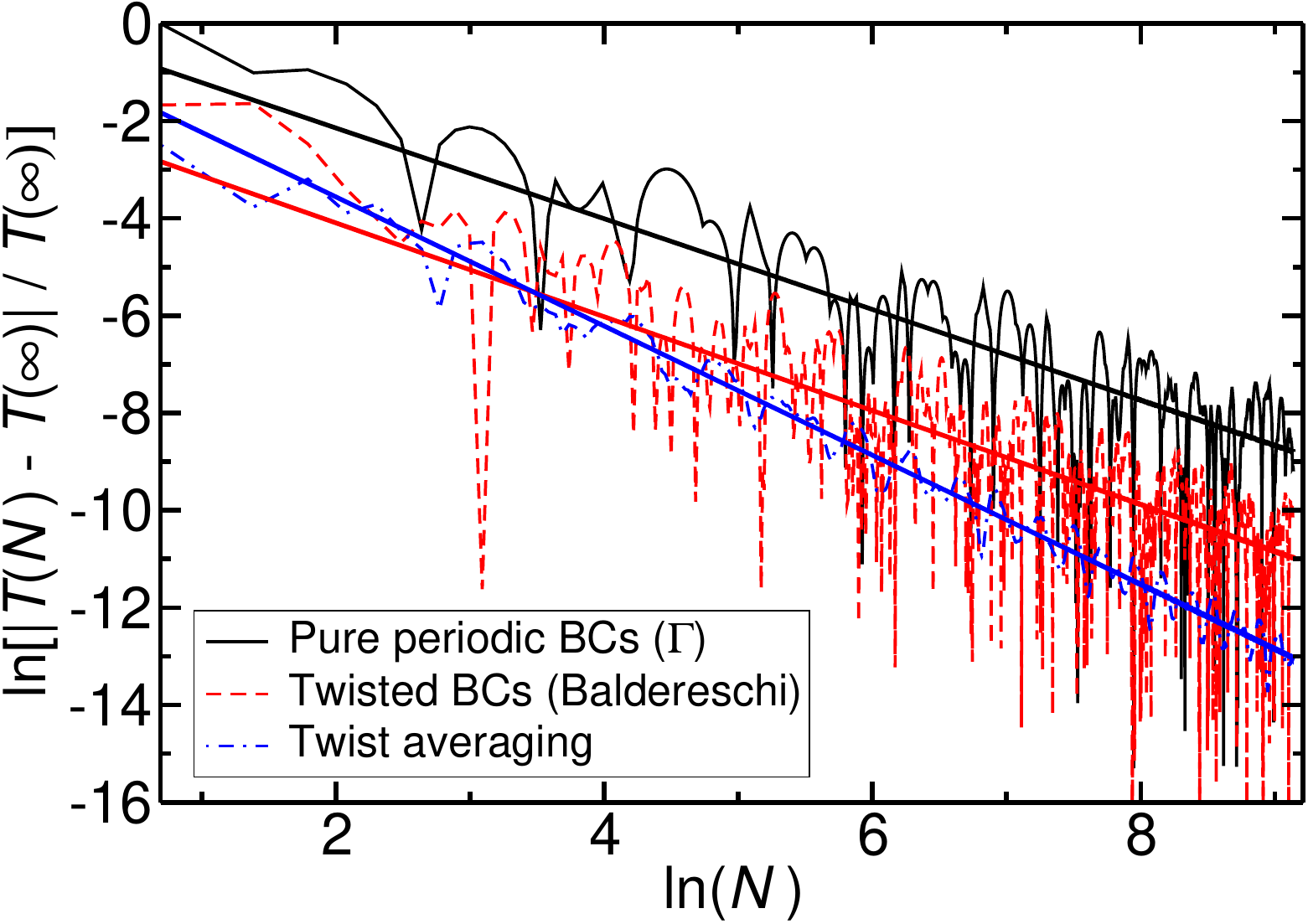}
\caption{\label{fig:HEG3D_SC_para_HF_T_v_N_log-log} Logarithm of
  absolute difference between kinetic energy in a finite cell and in
  an infinite system, divided by infinite-system kinetic energy,
  against logarithm of system size, for a two-component free electron
  gas in a simple cubic cell. Results are shown for periodic boundary
  conditions, twisted periodic boundary conditions at the
  simulation-cell Baldereschi point, and twist-averaged boundary
  conditions.  The solid lines show straight-line fits to the log-log
  plots; the gradients are $-1$ for the results with periodic and
  twisted periodic boundary conditions, and $-4/3$ for twist-averaged
  boundary conditions.}
\end{figure}

Figure \ref{fig:HEG3D_SC_para_HF_T_v_N_log-log} confirms that
finite-size errors in the twist-averaged kinetic energy per electron
fall off as $N^{-4/3}$ \cite{Lin_2001}, whereas the typical
finite-size errors with periodic ($\Gamma$) and twisted periodic
(Baldereschi) points fall off as $N^{-1}$, despite the fact that the
latter has a much smaller prefactor.  Hence, although the Baldereschi
point is clearly vastly preferable to the $\Gamma$ point in a finite
cell subject to twisted periodic boundary conditions, it is essential
to apply twist averaging in a metallic system.

Momentum quantisation finite-size errors in QMC simulations of
metallic systems such as interacting homogeneous electron gases behave
in a similar manner to the finite-size errors in the free electron
kinetic energy, except that the kinetic energies are slightly rescaled
by the quasiparticle effective mass according to Landau's Fermi liquid
theory. Twist-averaging (especially with further corrections for
residual errors in twist-averaged energies \cite{Drummond_2008})
therefore remains a crucial method for removing momentum quantisation
errors in metallic systems.  However, wherever there is a task that is
to be performed at one particular set of twisted boundary conditions,
such as optimising a Jastrow factor or backflow function to be used at
all twists, the Baldereschi point is in general a very much more
representative twist than $\Gamma$, even in a metallic system.

\section{Conclusions}

We have evaluated and tabulated Baldereschi mean-value points for all
fourteen Bravais lattices in 3D\@.  We expect that these results will
be particularly useful in QMC studies of condensed matter.

\ack{Our graphene calculations were performed using Lancaster
  University's High End Computing (HEC) cluster.}

%\funding{}

%\roles{}

%\data{}

\appendix

\section{Algorithm for identifying the target stars for a given
Bravais lattice\label{app:targets}}

We identify a list of target star functions to be set to zero or
minimised, in order of priority, as follows.
\begin{enumerate}

\item Calculate the set of lattice points $\{{\bf R}\}$ and sort them
  into the star structure, with the stars $\bigstar_n$ in ascending
  order of radius $r_n$, with $\bigstar_0=\{{\bf 0}\}$.

\item Let our first target star be $t_1=1$.

\item If $\dim(\spn(\bigstar_{t_1}))<3$ then search for the lowest
  $t_2>t_1$ such that $\dim(\spn(\bigstar_{t_1} \cup \bigstar_{t_2}))
  > \dim(\bigstar_{t_1})$; if $\dim(\spn(\bigstar_{t_1} \cup
  \bigstar_{t_2}))<3$ then search for the lowest $t_3>t_2$ such that
  $\dim(\spn(\bigstar_{t_1} \cup \bigstar_{t_2} \cup
  \bigstar_{t_3}))=3$.

\item Loop over stars $n$:
\begin{enumerate}

\item Let $T_n$ be the set of targets $t_i$ such that $t_i<n$ and
  $\spn(\bigstar_{t_i}) \cap \spn(\bigstar_n) \neq \{ {\bf 0} \}$.

\item If $(2r_n)^2 \neq \sum_{t \in T_n} (n_t r_t)^2$ for all sets of
  integers $\{n_t\}$ then choose $t=n$ to be the next target star in
  our list.

\end{enumerate}

\item Loop over targets $t_i$ in order of increasing radius:
\begin{enumerate}

\item Test whether $(2r_{t_i})^2=\sum_{j \neq i} (n_j r_{t_j})^2$ for
  $n_j \in \mathbb{Z}$; if so, identify the largest-radius target
  $t_j$ with nonzero $n_j$ that is not already listed for removal; add
  this target to a list of targets to remove.

\end{enumerate}

\item Reduce the list of targets by removing the redundant targets
  identified in the previous step.

\end{enumerate}

In practice steps 5 and 6 are unnecessary for 3D Bravais lattices.

\section{Algorithm for numerically determining the
Baldereschi point\label{app:numerics}}

Our algorithm for numerically zeroing/minimising the target star
functions (identified as described in appendix \ref{app:targets}) is as
follows.

\begin{enumerate}

\item Repeatedly:
\begin{enumerate}

\item Choose initial wavevector ${\bf k}$ using a Sobol' sequence
  with a constant offset.

\item Try to solve $A_{t_1}({\bf k})=A_{t_2}({\bf k})=A_{t_3}({\bf
  k})=0$ using Newton-Raphson iteration starting from the sampled
  ${\bf k}$.  If that succeeds and $A_{t_4}$ is available then record
  the value of $g=A_{t_4}^2({\bf k})$ at the solution.  Otherwise:
\begin{enumerate}

\item Try to solve $A_{t_1}({\bf k})=A_{t_2}({\bf k})=0$ using Newton-Raphson
  iteration starting from the sampled ${\bf k}$.  If that succeeds then:
\begin{enumerate}

\item minimise $A_{t_3}^2({\bf k})+\left[ A_{t_1}^2({\bf k}) + A_{t_2}^2({\bf k})
  \right]/\sqrt{\epsilon}$ using BFGS;

\item solve $A_{t_1}({\bf k})=A_{t_2}({\bf k})=0$ using Newton-Raphson iteration;

\item find the constrained stationary point of $A_{t_3}({\bf k})$ subject
  to $A_{t_1}({\bf k})=A_{t_2}({\bf k})=0$ using Lagrange multipliers and
  Newton-Raphson iteration;

\item record the value of $g=A_{t_3}^2({\bf k})$;

\end{enumerate}
otherwise:
\begin{enumerate}
\setcounter{enumiv}{4}

\item solve $A_{t_1}({\bf k})=0$ using Newton-Raphson iteration starting from
  the sampled ${\bf k}$;

\item minimise $A_{t_2}^2({\bf k})+A_{t_1}^2({\bf k})/\sqrt{\epsilon}$
  using BFGS;

\item solve $A_{t_1}({\bf k})=0$ using Newton-Raphson iteration;

\item find the constrained stationary point of $A_{t_2}({\bf k})$ subject
  to $A_{t_1}({\bf k})=0$ using Lagrange multipliers and Newton-Raphson
  iteration;

\item record the value of $g=A_{t_2}^2({\bf k})$.

\end{enumerate}

\end{enumerate}

\end{enumerate}

\item Identify the Baldereschi point as the solution for which,
  firstly, the number of target star functions set to zero is
  maximised and, secondly, the value of $g$ is minimised.

\end{enumerate}

\bibliographystyle{iopart-num} \bibliography{baldereschi}

@article{Baldereschi_1973,
title={Mean-Value Point in the {B}rillouin Zone},
author={Baldereschi, A.},
journal={Phys. Rev. B},
volume={7},
issue={12},
pages={5212-5215},
numpages={0},
year={1973},
month={June},
publisher={American Physical Society},
doi={10.1103/PhysRevB.7.5212},
url={https://link.aps.org/doi/10.1103/PhysRevB.7.5212}}

@article{Chadi_1973,
title={Special Points in the {B}rillouin Zone},
author={Chadi, D. J. and Cohen, Marvin L.},
journal={Phys. Rev. B},
volume={8},
issue={12},
pages={5747-5753},
numpages={0},
year={1973},
month={December},
publisher={American Physical Society},
doi={10.1103/PhysRevB.8.5747},
url={https://link.aps.org/doi/10.1103/PhysRevB.8.5747}}

@article{Cunningham_1974,
title={Special points in the two-dimensional {B}rillouin zone},
author={Cunningham, S. L.},
journal={Phys. Rev. B},
volume={10},
issue={12},
pages={4988-4994},
numpages={0},
year={1974},
month={December},
publisher={American Physical Society},
doi={10.1103/PhysRevB.10.4988},
url={https://link.aps.org/doi/10.1103/PhysRevB.10.4988}}

@article{Monkhorst_1976,
title={Special points for {B}rillouin-zone integrations},
author={Monkhorst, Hendrik J. and Pack, James D.},
journal={Phys. Rev. B},
volume={13},
issue={12},
pages={5188-5192},
numpages={0},
year={1976},
month={June},
publisher={American Physical Society},
doi={10.1103/PhysRevB.13.5188},
url={https://link.aps.org/doi/10.1103/PhysRevB.13.5188}}

@article{Bashenov_1977,
author={Bashenov, V. K. and Bardashova, M. and Mutal, A. M.},
title={Baldereschi Points for Some Noncubic Lattices},
journal={Phys. Status Solidi B},
volume={80},
number={2},
pages={K89-K93},
doi={https://doi.org/10.1002/pssb.2220800239},
url={https://onlinelibrary.wiley.com/doi/abs/10.1002/pssb.2220800239},
year={1977}}

@article{Chulkov_1979,
author={Chulkov, E. V.
and Psakh'e, S. G.},
title={Special points in fct structures},
journal={Sov. Phys. J.},
year={1979},
month={November},
day={01},
volume={22},
number={11},
pages={1237-1238},
issn={1573-9228},
doi={10.1007/BF00894985},
url={https://doi.org/10.1007/BF00894985}}

@article{Evarestov_1983,
author={Evarestov, R. A. and Smirnov, V. P.},
title={Special points of the {B}rillouin zone and their use in the solid state theory},
journal={Phys. Status Solidi B},
volume={119},
number={1},
pages={9-40},
doi={https://doi.org/10.1002/pssb.2221190102},
url={https://onlinelibrary.wiley.com/doi/abs/10.1002/pssb.2221190102},
year={1983}}

@misc{Stevanovic_2024,
title={A {P}ython code for calculating the mean-value
({B}aldereschi's) point for any crystal structure}, 
author={Vladan Stevanovic},
year={2024},
archivePrefix={arXiv},
primaryClass={cond-mat.other},
url={https://arxiv.org/abs/2405.00925}}

@article{Rajagopal_1994,
title={Quantum {M}onte {C}arlo Calculations for Solids Using Special $k$ Points Methods},
author={Rajagopal, G. and Needs, R. J. and Kenny, S. and Foulkes, W. M. C. and James, A.},
journal={Phys. Rev. Lett.},
volume={73},
issue={14},
pages={1959-1962},
numpages={0},
year={1994},
month={October},
publisher={American Physical Society},
doi={10.1103/PhysRevLett.73.1959},
url={https://link.aps.org/doi/10.1103/PhysRevLett.73.1959}}

@article{Rajagopal_1995,
title={Variational and diffusion quantum {M}onte {C}arlo calculations at nonzero wave vectors: Theory and application to diamond-structure germanium},
author={Rajagopal, G. and Needs, R. J. and James, A. and Kenny, S. D. and Foulkes, W. M. C.},
journal={Phys. Rev. B},
volume={51},
issue={16},
pages={10591-10600},
numpages={0},
year={1995},
month={April},
publisher={American Physical Society},
doi={10.1103/PhysRevB.51.10591},
url={https://link.aps.org/doi/10.1103/PhysRevB.51.10591}}

@article{Drummond_2008,
title={Finite-size errors in continuum quantum {M}onte {C}arlo calculations},
author={Drummond, N. D. and Needs, R. J. and Sorouri, A. and Foulkes, W. M. C.},
journal={Phys. Rev. B},
volume={78},
issue={12},
pages={125106},
numpages={19},
year={2008},
month={September},
publisher={American Physical Society},
doi={10.1103/PhysRevB.78.125106},
url={https://link.aps.org/doi/10.1103/PhysRevB.78.125106}}

@article{Azadi_2015,
author={Azadi, Sam and Foulkes, W. M. C.},
title={Systematic study of finite-size effects in quantum {M}onte {C}arlo calculations of real metallic systems},
journal={J. Chem. Phys.},
volume={143},
number={10},
pages={102807},
year={2015},
month={June},
issn={0021-9606},
doi={10.1063/1.4922619},
url={https://doi.org/10.1063/1.4922619}}

@article{Lin_2001,
title={Twist-averaged boundary conditions in continuum quantum {M}onte {C}arlo algorithms},
author={Lin, C. and Zong, F. H. and Ceperley, D. M.},
journal={Phys. Rev. E},
volume={64},
issue={1},
pages={016702},
numpages={12},
year={2001},
month={June},
publisher={American Physical Society},
doi={10.1103/PhysRevE.64.016702},
url={https://link.aps.org/doi/10.1103/PhysRevE.64.016702}}

@article{Sobol_1976,
title={Uniformly distributed sequences with an additional uniform property},
journal={USSR Comp. Math. Math. Phys.},
volume={16},
number={5},
pages={236-242},
year={1976},
issn={0041-5553},
doi={https://doi.org/10.1016/0041-5553(76)90154-3},
url={https://www.sciencedirect.com/science/article/pii/0041555376901543},
author={I.M. Sobol}}

@article{Bratley_1988,
author={Bratley, Paul and Fox, Bennett L.},
title={Algorithm 659: Implementing {S}obol's quasirandom sequence generator},
year={1988},
issue_date={March 1988},
publisher={Association for Computing Machinery},
address={New York, NY, USA},
volume={14},
number={1},
issn={0098-3500},
url={https://doi.org/10.1145/42288.214372},
doi={10.1145/42288.214372},
journal={ACM Trans. Math. Softw.},
month=mar,
pages={88-100},
numpages={13}}

@article{Helfrich_1985,
title={Algorithms to construct {M}inkowski reduced and {H}ermite reduced lattice bases},
journal={Theor. Comp. Sci.},
volume={41},
pages={125-139},
year={1985},
issn={0304-3975},
doi={https://doi.org/10.1016/0304-3975(85)90067-2},
url={https://www.sciencedirect.com/science/article/pii/0304397585900672},
author={Bettina Helfrich}}

@article{Fraser_1996,
title={Finite-size effects and {C}oulomb interactions in quantum
{M}onte {C}arlo calculations for homogeneous systems
with periodic boundary conditions},
author={Fraser, Louisa M. and Foulkes, W. M. C. and Rajagopal, G. and
Needs, R. J. and Kenny, S. D. and Williamson,
A. J.},
journal={Phys. Rev. B},
volume={53},
issue={4},
pages={1814-1832},
numpages={0},
year={1996},
month={January},
publisher={American Physical Society},
doi={10.1103/PhysRevB.53.1814},
url={https://link.aps.org/doi/10.1103/PhysRevB.53.1814}}

@article{Chiesa_2006,
title={Finite-Size Error in Many-Body Simulations with Long-Range
Interactions},
author={Chiesa, Simone and Ceperley, David M. and Martin, Richard
M. and Holzmann, Markus},
journal={Phys. Rev. Lett.},
volume={97},
issue={7},
pages={076404},
numpages={4},
year={2006},
month={August},
publisher={American Physical Society},
doi={10.1103/PhysRevLett.97.076404},
url={https://link.aps.org/doi/10.1103/PhysRevLett.97.076404}}

@article{Needs_2020,
author={Needs, R. J. and Towler, M. D. and Drummond, N. D. and L{\'o}pez R{\'i}os, P. and Trail, J. R.},
title={Variational and diffusion quantum {M}onte {C}arlo calculations with the {CASINO} code},
journal={J. Chem. Phys.},
volume={152},
number={15},
pages={154106},
year={2020},
month={April},
issn={0021-9606},
doi={10.1063/1.5144288},
url={https://doi.org/10.1063/1.5144288}}

@misc{CASINO,
author={Needs, R. J. and Towler, M. D. and Drummond, N. D. and
                  L\'{o}pez R\'{\i}os, P.},
title={\textsc{Casino} quantum {M}onte {C}arlo software.},
year={2026},
url={https://vallico.net/casinoqmc/},
note={{T}he \textsc{casino} program is available to download for
                  academic use, worldwide, free of cost.  For
                  commercial use of the software, please discuss with
                  the authors.  Access to \textsc{casino}'s Git
                  repository is available for academic use on request
                  to the authors. The program for calculating
                  Baldereschi points is available as part of the
                  \textsc{kvec\_maker} utility.}}

@article{Clark_2005, 
author={S. J. Clark and M. D. Segall and C. J. Pickard and
P. J. Hasnip and M. I. J. Probert and K. Refson and M. C. Payne},
title={First principles methods using {CASTEP}},
year={2005},
month={2014},
day={09-10T00:27:17.837+02:00},
journal={Z. Kristallogr.},
volume={220},
pages={567-570},
doi={10.1524/zkri.220.5.567.65075},
url={http://www.degruyter.com/view/j/zkri.2005.220.issue-5-6-2005/zkri.220.5.567.65075/zkri.220.5.567.65075.xml}}

@article{Lynch_1966,
author={Lynch, R. W. and Drickamer, H. G.},
title={Effect of High Pressure on the Lattice Parameters of Diamond, Graphite, and Hexagonal Boron Nitride},
journal={J. Chem. Phys.},
volume={44},
number={1},
pages={181-184},
year={1966},
month={January},
issn={0021-9606},
doi={10.1063/1.1726442},
url={https://doi.org/10.1063/1.1726442}}

@article{Drummond_2004,
title={{J}astrow Correlation Factor for Atoms, Molecules, and
Solids},
author={Drummond, N. D. and Towler, M. D. and Needs, R. J.},
journal={Phys. Rev. B},
volume={70},
issue={23},
pages={235119},
numpages={11},
year={2004},
month={December},
publisher={American Physical Society},
doi={10.1103/PhysRevB.70.235119},
url={https://link.aps.org/doi/10.1103/PhysRevB.70.235119}}

@article{Trail_2005a,
author={Trail, J. R. and Needs, R. J.},
title={Norm-Conserving {H}artree-{F}ock Pseudopotentials and
Their Asymptotic Behavior},
journal={J. Chem. Phys.},
year={2005},
volume={122},
number={1}, 
pages=014112,
url=
{http://scitation.aip.org/content/aip/journal/jcp/122/1/10.1063/1.1829049},
doi={http://dx.doi.org/10.1063/1.1829049}}

@article{Trail_2005b,
author={Trail, J. R. and Needs, R. J.},
title={Smooth Relativistic {H}artree-{F}ock Pseudopotentials for
{H} to {B}a and {L}u to {H}g},
journal={J. Chem. Phys.},
year={2005},
volume={122},
number={17},
pages=174109,
url=
{http://scitation.aip.org/content/aip/journal/jcp/122/17/10.1063/1.1888569},
doi={http://dx.doi.org/10.1063/1.1888569}}

@article{Umrigar_2007,
title={Alleviation of the Fermion-Sign Problem by Optimization of Many-Body
Wave Functions},
author={Umrigar, C. J. and Toulouse, J. and Filippi, C. and
Sorella, S. and Hennig, R. G.},
journal={Phys. Rev. Lett.},
volume={98},
issue={11},
pages={110201},
numpages={4},
year={2007},
month={March},
publisher={APS},
doi={10.1103/PhysRevLett.98.110201},
url={http://link.aps.org/doi/10.1103/PhysRevLett.98.110201}}

\end{document}